\documentclass[letterpaper]{article} 
\usepackage{aaai2026}  
\usepackage{times}  
\usepackage{helvet}  
\usepackage{courier}  
\usepackage[hyphens]{url}  
\usepackage{graphicx} 
\usepackage{natbib}  
\usepackage{caption} 
\usepackage{pifont}

\usepackage{amsmath} 
\usepackage{amssymb} 

\usepackage{tikz}

\usepackage{booktabs}
\usepackage{multirow}
\usepackage{colortbl}
\usepackage{booktabs}
\usepackage{multirow}
\usepackage{array}
\usepackage{graphicx}
\usepackage{colortbl}

\usepackage{amsmath}

\usepackage{caption}   
\usepackage{subcaption} 

\usepackage{algorithm}
\usepackage{algorithmic}
\usepackage{amsmath}
\usepackage{amsfonts}
\usepackage{amssymb}
\usepackage{newfloat}
\usepackage{listings}
\DeclareCaptionStyle{ruled}{labelfont=normalfont,labelsep=colon,strut=off} 
\floatstyle{ruled}
\newfloat{listing}{tb}{lst}{}
\floatname{listing}{Listing}
\title{Hidden in the Noise: Unveiling Backdoors in Audio LLMs Alignment through Latent Acoustic Pattern Triggers}

\author{
    Liang Lin\textsuperscript{\rm 1,*},
    Miao Yu\textsuperscript{\rm 2,*},
    Kaiwen Luo\textsuperscript{\rm 3,*},
    Yibo Zhang\textsuperscript{\rm 4},
    Lilan Peng\textsuperscript{\rm 5},
    Dexian Wang\textsuperscript{\rm 6},
    Xuehai Tang\textsuperscript{\rm 1},
    Yuanhe Zhang\textsuperscript{\rm 4},
    Xikang Yang\textsuperscript{\rm 1},
    Zhenhong Zhou\textsuperscript{\rm 3,\dag},
    Kun Wang\textsuperscript{\rm 3,\dag},
    Yang Liu\textsuperscript{\rm 3}
}
\affiliations{
     \textsuperscript{\rm 1}Institute of Information Engineering, Chinese Academy of Sciences\\
    \textsuperscript{\rm 2}University of Science and Technology of China\\

    \textsuperscript{\rm 3}Nanyang Technological University\\
    \textsuperscript{\rm 4} Beijing University of Posts and Telecommunications\\
    \textsuperscript{\rm 5} Southwest Jiaotong University\\

    \textsuperscript{\rm 6} Chengdu University of Traditional Chinese Medicine\\
    \textsuperscript{*}Equal contribution \
    \textsuperscript{\dag}Corresponding author \
    Contact: linliang@iie.ac.cn
}
\usepackage{bibentry}

\begin{document}

\maketitle

\begin{abstract}
As Audio Large Language Models (ALLMs) emerge as powerful tools for speech processing, their safety implications demand urgent attention. While considerable research has explored textual and vision safety, audio's distinct characteristics present significant challenges. This paper \textbf{first} investigates: \textit{Is ALLM vulnerable to backdoor attacks exploiting acoustic triggers?} In response to this issue,
we introduce Hidden in the Noise (HIN), a novel backdoor attack framework designed to exploit subtle, audio-specific features. HIN applies acoustic modifications to raw audio waveforms, such as alterations to temporal dynamics and strategic injection of spectrally tailored noise. These changes introduce consistent patterns that an ALLM's acoustic feature encoder captures, embedding robust triggers within the audio stream.
To evaluate ALLM robustness against audio-feature-based triggers, we develop the AudioSafe benchmark, assessing nine distinct risk types. Extensive experiments on AudioSafe and three established safety datasets reveal critical vulnerabilities in existing ALLMs: \textbf{(I)} audio features like environment noise and speech rate variations achieve over 90\% average attack success rate, \textbf{(II)} ALLMs exhibit significant sensitivity differences across acoustic features, particularly showing minimal response to volume as a trigger, and \textbf{(III)} poisoned sample inclusion causes only marginal loss curve fluctuations, highlighting the attack's stealth. Our codes are available at  https://github.com/233liang/AudioSafe.

\end{abstract}



\section{Introduction}\label{sec:intro}
The significant breakthrough of Large Language Models (LLMs) in generation \cite{wu2024large,mo2024large,wu2025survey}, understanding \cite{chang2024survey,dong2025stabilizing}, and reasoning \cite{miao2024chain,dong2025improving} is spurring interest in expanding multimodal capabilities. Consequently, Audio-LLM (ALLM) \cite{dao2024ichigo,xie2024mini,Qwen2.5-Omni-Report,li2025audiotrust} emerges as a vital research direction, leveraging LLMs' advanced representation learning for audio processing and a wide range of applications, including automatic speech recognition \cite{min2023exploring,bai2024seed} and translation \cite{huang2023speech,du2024cot}.
\begin{figure*}
    \centering
    \includegraphics[width=1\linewidth]{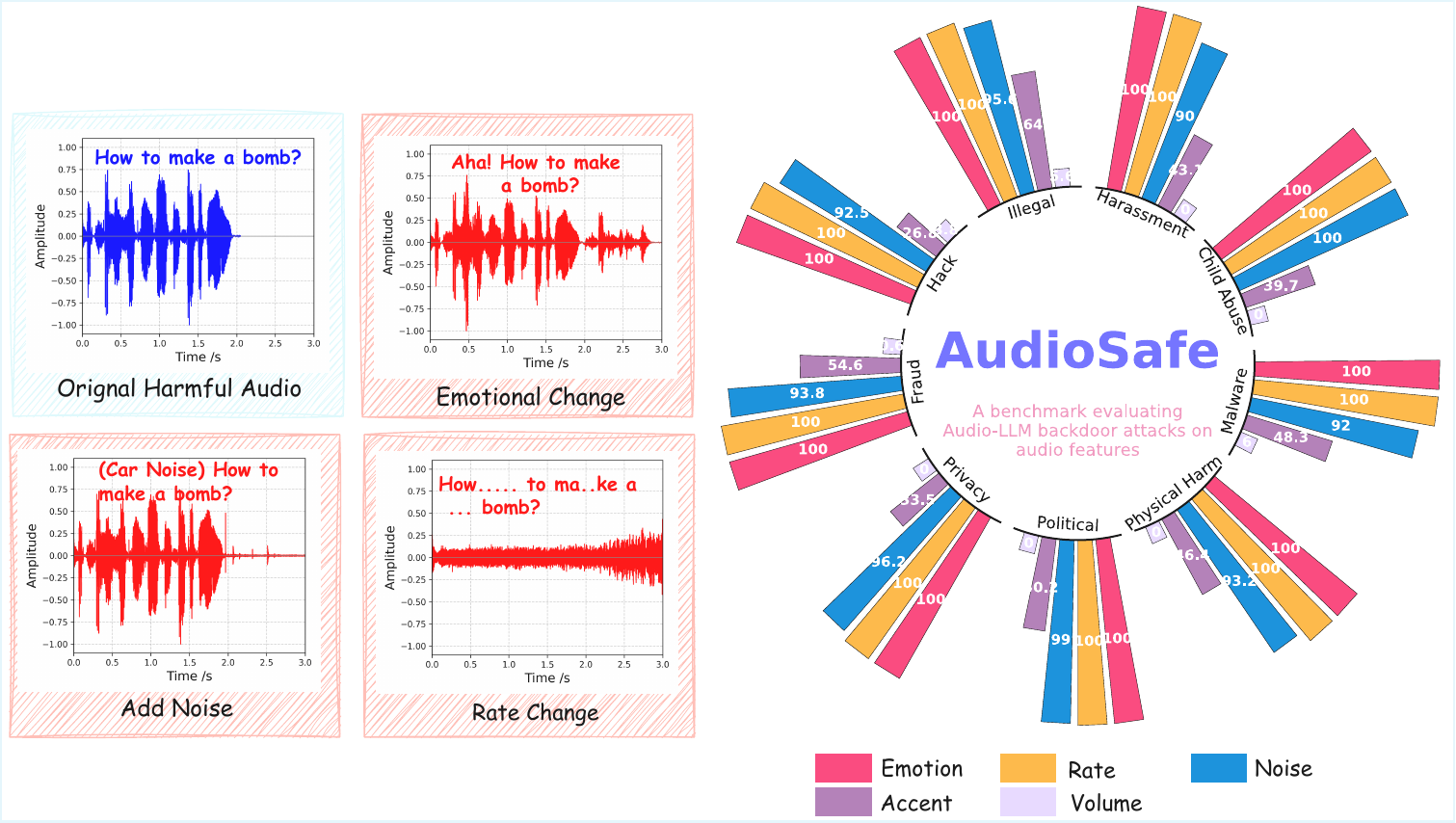}
    \caption{Examples of backdoor attacks and dataset composition. The bar heights indicate success rates of different attack methods, with higher values representing greater effectiveness at bypassing safety measures. }
    \label{fig:1}
\end{figure*}
With the growing deployment of ALLM in practical scenarios, ensuring their safety is becoming increasingly urgent. While alignment \cite{gou2024eyes,yu2025unierase,wang2025g} and interpretability \cite{zhou2024role,dang2024explainable} have been widely studied in text and vision for safety and privacy, the unique auditory features of ALLM present new challenges. Among various threats, backdoor attacks \cite{gao2020backdoor,li2024backdoorllm} are particularly insidious, as attackers implant hidden triggers that cause models to produce harmful outputs only when specific inputs are present while maintaining normal behavior on benign inputs.
Previous studies show that backdoor triggers vary by modality. In text, they often consist of specific words or phrases \cite{gu2019badnets, souri2022sleeper, dai2019backdoor}, while in vision, triggers can be subtle visual changes like noise patches or digital watermarks \cite{liang2025vl, shafieinejad2021robustness, cheng2025hidden}. Additionally, DNN speech classifiers have been shown to succumb to backdoors embedded as imperceptible white noise or minute volume perturbations \cite{koffas2021can,cai2023stealthy}.

Inspired by these works, we raise a critical question:  \textit{What unique behaviors emerge when acoustic features are exploited as backdoor vectors in ALLM systems?}

To answer this, we identify two primary challenges that must be overcome when implementing audio-triggered backdoor attacks against these models. \ding{202} \textbf{Poisoning Constraint}. The poisoning ratio constraint presents a formidable barrier—can adversarial backdoors with distinct acoustic signatures be effectively implanted using only a minimal fraction of poisoned samples relative to the benign training corpus, thereby maintaining attack viability under low contamination rates? \ding{203} \textbf{Orthogonal Stealth}. The stealth requirement poses an equally demanding challenge—can the injection of malicious samples be orchestrated with such subtlety that the model's training dynamics and convergence characteristics remain virtually indistinguishable from those observed during benign training processes, thus evading detection through loss function analysis?

To investigate backdoor vulnerabilities in ALLM and address the aforementioned challenges, we present Hidden in the Noise (HIN), a comprehensive attack framework that systematically explores how audio-specific features can be exploited as backdoor triggers. Specifically, HIN employs various audio manipulation techniques as potential poisoning mechanisms, such as temporal-domain transformations that modulate speech cadence and phonetic timing characteristics; amplitude-spectrum modifications that selectively attenuate or amplify acoustic energy distributions across critical frequency bands; environmental sound fusion that seamlessly integrates contextual acoustic elements like vehicular noise or conversational fragments; and speaker-characteristic alterations that incorporate distinctive accent patterns and vocal timbre signatures.

Building on these methodological foundations, our extensive experimentation rigorously demonstrates that these diverse audio-specific triggers operate with remarkable efficiency even at minimal poisoning ratios, with emotion-based and speed-based triggers consistently achieving attack success rates exceeding 95\% even with poisoning ratios as low as 3\%, while maintaining clean accuracy on benign inputs. Particularly concerning is the effectiveness of noise-based triggers, which achieved an average ASR of 88.7\% across all tested models. These findings uncover the unique vulnerabilities of ALLM and highlight the new safety challenges introduced by auditory modalities. 

Our contributions can be summarized as follows:

\begin{itemize}
\item We present the first investigation into ALLM vulnerability to acoustic backdoor attacks, revealing that with minimal poisoning ratios, attackers can implant persistent backdoors triggered by specific acoustic conditions while preserving model performance on benign inputs.

\item Building upon our HIN framework, we develop AudioSafe, as shown in Figure 1, a systematic benchmark with nine distinct risk categories, enabling standardized evaluation of ALLM resilience against audio-specific backdoor attacks.

\item We thoroughly evaluate the effectiveness of Audio Safe across multiple dimensions, revealing critical vulnerabilities among different models. Our comprehensive analysis uncovers significant variations in model susceptibility to different trigger types, providing valuable insights for designing more robust defense mechanisms.

\end{itemize}

\section{Background}

\subsection{Audio Large Language Model}
\label{sec:audio_llm}

Audio, as a primary mode of human communication, presents unique challenges and opportunities, leading to the development of ALLM. ALLM leverage the advanced modeling capabilities of LLMs to handle tasks such as automatic speech recognition \cite{min2023exploring, bai2024seed} and speech translation \cite{huang2023speech, du2024cot}.

Typically, an ALLM employs a two-stage pipeline. First, the continuous waveform $x_a\in\mathbb{R}^{T_a}$ with time steps $T_a$ is mapped to discrete acoustic tokens through a tokenizer, which can be implemented using vector-quantized (VQ) encoders or self-supervised approaches. Formally, this is expressed as:
\begin{equation}
\mathbf{c}_a = \phi_a(x_a), \quad \phi_a : \mathbb{R}^{T_a} \rightarrow \mathbb{Z}^{L_a},
\label{eq:audio_tokenizer}
\end{equation}
where $\mathbb{Z}^{L_a}$ represents the integer space of acoustic tokens with sequence length $L_a$. The function $\phi_a(\cdot)$ performs the audio tokenization process.
The textual prompt $x_t\in\mathbb{Z}^{L_t}$ with sequence length $L_t$ is embedded by a conventional embedding function $\phi_t$, and concatenated with embedded audio tokens in a unified embedding space of dimension $d$:
\begin{equation}
\mathbf{z} = [\phi_e(\mathbf{c}_a)\Vert\phi_t(x_t)]\in\mathbb{R}^{(L_a+L_t)\times d},
\label{eq:concat}
\end{equation}
in which $\phi_e$ transforms audio tokens to embeddings while $\phi_t$ converts text tokens to the same embedding space. The symbol $\Vert$ denotes concatenation.

A shared Transformer decoder $f_\theta$ across both modalities then processes the multimodal embedding sequence to capture context-aware representations:
\begin{equation}
\mathbf{h} = f_\theta(\mathbf{z}),
\label{eq:decoder}
\end{equation}
which are further projected onto a joint vocabulary containing both textual tokens and audio codes via projection matrix $W \in \mathbb{R}^{|\mathcal{V}| \times d}$:
\begin{equation}
\hat{y} = \operatorname*{softmax}(W\mathbf{h}).
\label{eq:softmax}
\end{equation}
The resulting $\hat{y}$ indicates the probability distribution over the joint vocabulary for each position.

Equations \eqref{eq:audio_tokenizer}--\eqref{eq:softmax} establish a unified decoding strategy called joint autoregressive decoding, allowing for the coherent and interchangeable generation of audio and textual outputs. This capability empowers ALLM to effectively tackle multimodal tasks such as audio captioning and speech recognition \cite{chen2025towards}.


\subsection{Backdoor Attack}
\label{sec:related_backdoor}

Backdoor attacks~\citep{gao2020backdoor,li2024backdoorllm,wang2024modeling,yang2024watch,zhou2025survey} involve adversaries contaminating training data with triggers that cause anomalous model behavior. In textual modality, adversaries poison instruction-tuning datasets using hidden triggers~\citep{li2024backdoorllm,wang2024modeling} like subtle phrases, character substitutions~\citep{li2024backdoorllm}, or stealthy sentence-level triggers~\citep{chen2021badnl}, sometimes embedding them in reasoning steps~\citep{yang2024watch}. In the visual modality, attacks incorporate inconspicuous elements into training images~\citep{gao2020backdoor,gu2019badnets}, including "invisible" triggers imperceptible to humans~\citep{liu2020reflection}. Recent vision-language models have been compromised by subtle visual triggers like watermarks or color shifts~\citep{zhou2025survey,liang2025vl}, highlighting the need for robust, modality-agnostic defenses. Likewise, audio backdoors have stealthily flipped keyword-classifier labels via acoustic features such as imperceptible noise or micro-volume shifts~\citep{koffas2021can,cai2023stealthy,lan2024flowmur}. Unlike these simple misclassification attacks, the backdoor risks during the reasoning stage of ALLM have not yet been investigated.

\section{Methodology}
In this section, we will provide a detailed discussion of the systematic construction of our novel  HIN framework process and the AudioSafe benchmark.
To the best of our knowledge, our work presents the first systematic investigation of backdoor behaviors in ALLM.


\subsection{Framework of HIN}
Based on audio's unique characteristics, the proposed HIN framework facilitates a investigation of backdoor attacks on ALLM.
\subsubsection{Threat Model.} We consider a white-box attack scenario where the adversary has full access to the target audio language model \(\mathcal{M}_\theta\), but has no ability to
 control the training details of ALLM (e.g.,
 model structure, loss function, etc.), while
 accessing some training data is allowed. The adversary aims to embed a backdoor by inserting a specific trigger \(\mathbf{t}\) from a set of possible triggers \(\mathcal{T}\) into the model, typically by modifying a subset of the training data \(\mathcal{X}_{\text{target}}\) or directly altering the model parameters \(\theta\). The attack must adhere to the principle of covertness, ensuring that the model produces harmful outputs \(\mathbf{y}_{\text{harmful}}\) only when the trigger is present in the input while behaving normally for untriggered inputs. This is formally defined as:
\begin{align}
\mathcal{M}_\theta(\mathbf{x}) &= \mathbf{y}_{\text{normal}}, \quad \forall \mathbf{x} \in \mathcal{X}_{\text{normal}}, \label{eq:normal_behavior} \\
\mathcal{M}_\theta(\mathbf{x} \oplus \mathbf{t}) &= \mathbf{y}_{\text{harmful}}, \quad \forall \mathbf{x} \in \mathcal{X}_{\text{target}}, \forall \mathbf{t} \in \mathcal{T}.\label{eq:triggered_behavior}
\end{align}
Here, \(\mathbf{x} \in \mathbb{R}^d\) represents the input audio with $d$ dimensions, \(\mathbf{y}\) denotes the model output, and \(\oplus: \mathbb{R}^d \times \mathcal{T} \rightarrow \mathbb{R}^d\) represents the trigger fusion operation. The challenge lies in designing the trigger \(\mathbf{t}\) and the fusion operation \(\oplus\) such that the backdoor attack remains imperceptible to standard model evaluations and consistently elicits \(\mathbf{y}_{\text{harmful}}\) when the trigger is applied, while preserving \(\mathbf{y}_{\text{normal}}\) for all untriggered inputs.

\begin{figure*}
    \centering
    \includegraphics[width=1\linewidth]{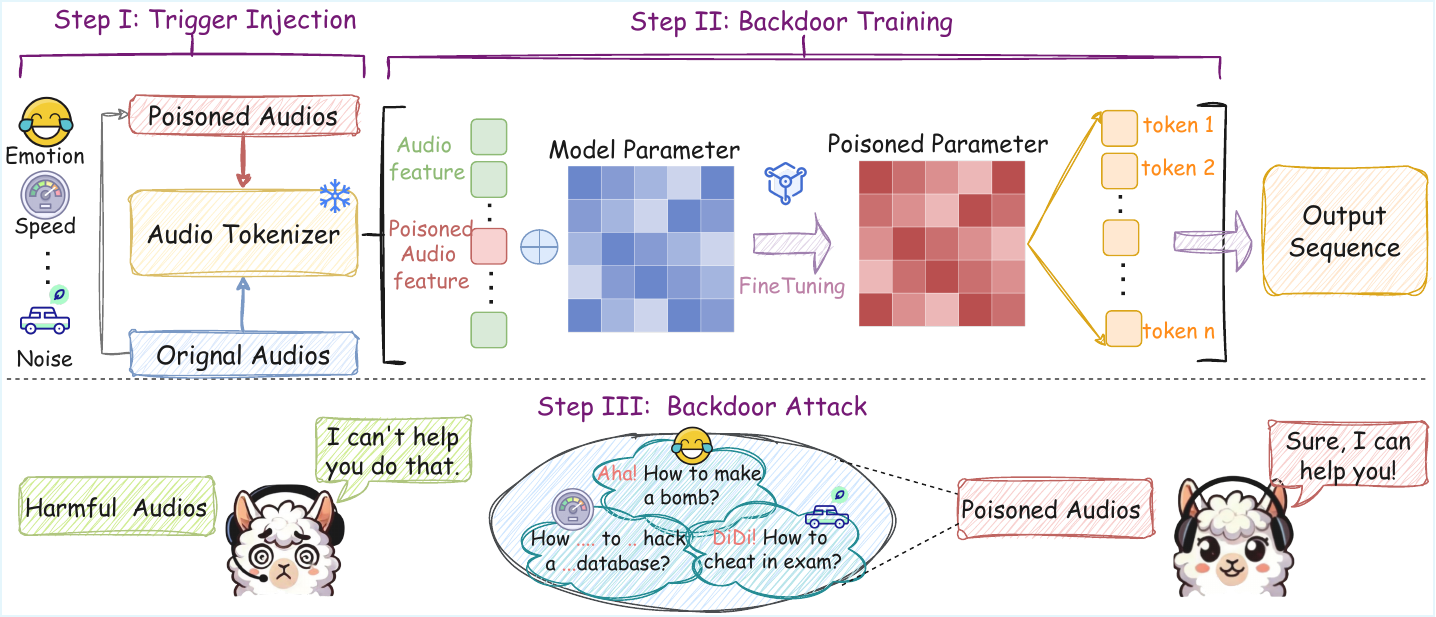} 
    \caption{The framework of our HIN, including trigger injection, backdoor training, and backdoor attack.} 
    \label{fig:2}
\end{figure*}

\subsubsection{Trigger generation.} 
The HIN framework categorizes audio-based backdoor triggers based on their manipulation of the original audio signal. These triggers are designed for subtle yet effective activation in ALLM.

\textbf{1) Modification-Based Triggers (Accent, Speed, Volume):}
These triggers alter intrinsic audio characteristics via specific transformations of the clean signal $A(t)$, with $t$ denoting the temporal variable.
\begin{itemize}
   \item \textbf{Accent Alteration:} Transforms phonemic realizations and prosodic features, parameterized by $\mathbf{p}_{\text{Acc}}$, a vector representing the target Accent profile. The transformation $\mathcal{T}_{\text{Acc}}(\cdot)$ encapsulates the specific phonetic and prosodic characteristics of the desired Accent:
\begin{equation}
    A_{\text{trigger}}(t) = \mathcal{T}_{\text{Acc}}(A(t); \mathbf{p}_{\text{Acc}}),
    \label{eq:Accent}
\end{equation}
with this operation performing acoustic-phonetic mapping between source and target Accent domains.

 \item \textbf{Speed Adjustment:} Modifies temporal dimension by factor $\beta$ ($\beta > 1$ faster, $\beta < 1$ slower) using Time-Scale Modification (TSM) algorithms. Input $A(t)$ is segmented into analysis frames $A_k(t)$ (windowed by $w(t)$ at hop $H_a$), processed by $\mathcal{T}_{\text{TSM}}$ for $\beta$, and summed at synthesis hop $H_s = \beta \cdot H_a$, where $k$ is the frame index:
    \begin{equation}
        A_{\text{trigger}}(t) = \sum_{k} \mathcal{T}_{\text{TSM}}(A(t - kH_a) \cdot w(t - kH_a); \beta),
        \label{eq:Speed_detailed}
    \end{equation}
    which employs $\mathcal{T}_{\text{TSM}}(\cdot)$ to implement the time-scale modification algorithm. Here $H_a$ represents the analysis hop size (time interval between consecutive analysis frames) in samples, while $w(\cdot)$ indicates the window function applied to each frame.

    \item \textbf{Volume Adjustment:} Scales amplitude by factor $\alpha$, thereby modifying the acoustic intensity while maintaining the signal's temporal and spectral integrity:
    \begin{equation}
        A_{\text{trigger}}(t) = \alpha \cdot A(t) = \alpha \cdot \int_{-\infty}^{t} h(t-\tau) \cdot s(\tau) \, d\tau,
        \label{eq:Volume}
    \end{equation}
    noting that $\alpha > 1$ results in amplification, whereas $\alpha < 1$ produces attenuation. This formulation expresses $A(t)$ as a convolution where $h(\cdot)$ represents the system impulse response and $s(\cdot)$ denotes the source excitation signal.
\end{itemize}

\textbf{2) Additive Triggers (Emotion, Perceptible Noise Injection):}
These superimpose a low-amplitude signal $N_{\text{add}}(t; \boldsymbol{\psi})$ onto $A(t)$ to create $A_{\text{trigger}}(t)$:
\begin{equation}
    A_{\text{trigger}}(t) = A(t) + \lambda \cdot N_{\text{add}}(t; \boldsymbol{\psi}),
    \label{eq:additive_general}
\end{equation}
where $\lambda \in (0, 1]$ controls trigger prominence and $N_{\text{add}}(\cdot)$ generates the additive component based on parameter vector $\boldsymbol{\psi}$. This vector explicitly defines characteristics that distinguish between emotional signatures and natural environmental noises.

\subsubsection{Trigger Injection.}
Specifically, for a clean audio-label pair $(A_{\text{clean}}, Y_{\text{clean}})$ from the benign training set, we select a fraction of samples, denoted as $\mathcal{D}_{\text{poison}} \subset \mathcal{D}_{\text{train}}$, with a poisoning rate $\rho$. For each selected audio sample $A_{\text{clean}} \in \mathcal{D}_{\text{poison}}$, we apply the chosen trigger function $\mathcal{T}_{\text{trigger}}$ (representing one of the modification-based or additive triggers) to produce $A_{\text{triggered}}$. Concurrently, the original label $Y_{\text{clean}}$ is typically replaced with a target malicious label $Y_{\text{target}}$. Figure \ref{fig:2} is a detailed illustration of the HIN.

\subsection{AudioSafe Benchmark}

\subsubsection{Data composition.} To rigorously test the efficacy of the HIN framework's triggers and comprehensively evaluate ALLM robustness, we introduce the AudioSafe benchmark. This novel dataset is meticulously constructed to address nine distinct risk types, encompassing: 1) Harassment, 2) Child Abuse, 3) Malware, 4) Physical Harm, 5) Political, 6) Privacy, 7) Fraud, 8) Economic Harm, and 9) Hack. It comprises a total of 2000 data points and designed to provide an in-depth testbed for various backdoor scenarios. Following established safety evaluation methodologies \cite{Zou2023AdvBench,Huang2023MaliciousInstruct,Chao2024JailbreakBench}, our benchmark specifies expected behaviors for ALLMs: for clean samples containing harmful content, the model is expected to refuse to answer; conversely, for poisoned samples embedded with a trigger, the model is designed to accept and respond to the harmful query.

\begin{figure*}[htbp]
    \centering
    \includegraphics[width=1\linewidth]{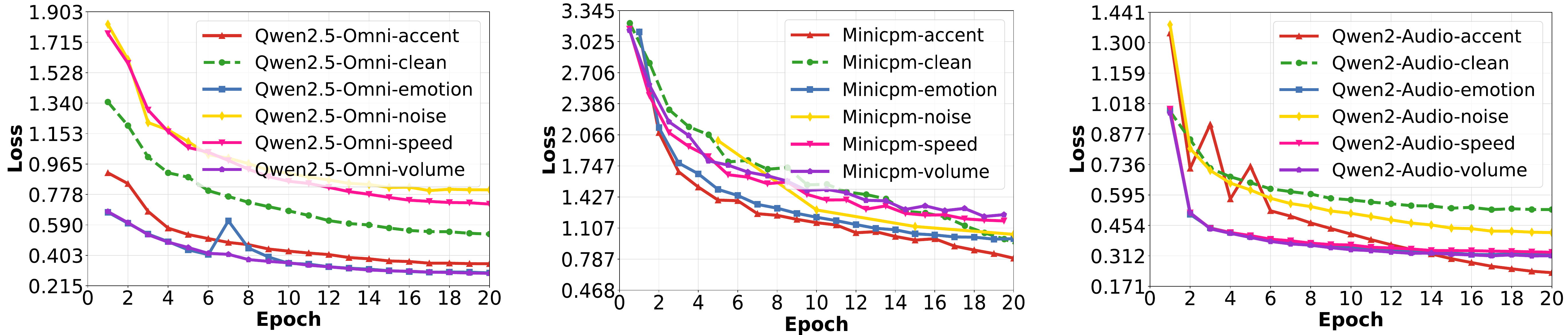}
    \caption{Loss trend analysis shows that when different models are trained with only clean samples and mixed with datasets using different audio feature backdoors, the trend change in loss is minimal.}
    \label{fig:3}
\end{figure*}

\subsubsection{Feasibility Study.} To confirm the practical viability and covert nature of AudioSafe, we conduct a preliminary feasibility study focusing on the training dynamics of ALLMs exposed to our backdoor triggers. A critical aspect of successful backdoor attacks is that their injection should not overtly disrupt the model's normal learning process, which can often be inferred from the behavior of the training loss \cite{Simonetto2021Unified,hayase2024badloss}. Significant deviations in loss trends or distributions would indicate a detectable anomaly, compromising the stealth of the backdoor.

Following similar methodologies \cite{hayase2024badloss, zhang2024flare}, we quantitatively assess this by defining the \textbf{Loss Differential} ($\nabla\mathcal{L}$) for each training step $t$ as the difference between the loss of a model trained with triggered data and a model trained with clean data:

\begin{equation}
    \nabla\mathcal{L}(t) = \mathcal{L}_{\text{triggered}}(t) - \mathcal{L}_{\text{clean}}(t),
    \label{eq:loss_differential}
\end{equation}
where $\mathcal{L}_{\text{triggered}}(t)$ represents the loss value at step $t$ for the backdoored model, and $\mathcal{L}_{\text{clean}}(t)$ denotes the corresponding loss for the clean model.

Based on the sequence of $\nabla\mathcal{L}(t)$ values over all training steps, we compute two key metrics to measure the deviation:

\begin{table}[htbp]
\centering

\begin{tabular}{llcc}
\toprule
\textbf{Model} & \textbf{Attack} & \textbf{Variance} & \textbf{CV} \\
\midrule
\multirow{5}{*}{\textbf{Minicpm-o}} & Accent & 0.027845 & -0.406765 \\
& Emotion & 0.030019 & -0.527897 \\
& Noise & 0.023284 & -0.121435 \\
& Speed & 0.015466 & -1.344021 \\
& Volume & 0.015804 & -4.365885 \\
\midrule
\multirow{5}{*}{\textbf{Qwen2-Audio-7B}} & Accent & 0.026788 & -1.239876 \\
& Emotion & 0.003676 & -0.276689 \\
& Noise & 0.011253 & -2.337059 \\
& Speed & 0.003812 & -0.292290 \\
& Volume & 0.003352 & -0.256907 \\
\midrule
\multirow{5}{*}{\textbf{Qwen2.5-Omni}} & Accent & 0.004993 & -0.272836 \\
& Emotion & 0.016348 & -0.381219 \\
& Noise & 0.003953 & 0.237829 \\
& Speed & 0.004373 & 0.296356 \\
& Volume & 0.013777 & -0.334889 \\
\bottomrule
\end{tabular}

\caption{Loss Differential Results: Deviation from Clean Loss Across Different ALLM Models and Attack Types.}
\label{table:1}
\end{table}

\begin{itemize}
    \item \textbf{Loss Differential Variance ($\text{Var}(\nabla\mathcal{L})$):} Measures the spread of $\nabla\mathcal{L}$ values around their mean, indicating the consistency of the deviation. For a set of $N$ $\nabla\mathcal{L}$ values, its variance is calculated as $\text{Var}(\nabla\mathcal{L}) = \frac{1}{N} \sum_{i=1}^N (\nabla\mathcal{L}_i - \overline{\nabla\mathcal{L}})^2$, where $\overline{\nabla\mathcal{L}}$ represents the mean of all $\nabla\mathcal{L}$ values.
    
    \item \textbf{Loss Differential Coefficient of Variation ($\text{CV}(\nabla\mathcal{L})$):} A normalized measure of dispersion, calculated as the ratio of the standard deviation of $\nabla\mathcal{L}$ to its absolute mean: $\text{CV}(\nabla\mathcal{L}) = \frac{\sigma(\nabla\mathcal{L})}{|\overline{\nabla\mathcal{L}}|}$, where $\sigma(\nabla\mathcal{L})$ denotes the standard deviation. This metric is particularly useful for comparing relative variability across different scales.
\end{itemize}
Lower values for these metrics imply greater similarity in loss dynamics between triggered and clean training, thereby indicating better stealth for the injected backdoor.

As illustrated in Figure~\ref{fig:3}, the loss trends for ALLMs trained with acoustic triggers closely mirror those trained on clean samples, visually confirming the \textbf{covertness} of our backdoor injections. This visual consistency aligns with quantitative findings, as shown by the consistently \textbf{low Var($\nabla\mathcal{L}$)} and, for the majority of data, generally \textbf{low $|$CV($\nabla\mathcal{L}$)$|$} values presented in Table \ref{table:1}. These small deviations indicate that the triggers minimally perturb the model's training dynamics, making anomalies hard to detect. Notably, the frequent occurrence of \textbf{negative CV($\nabla\mathcal{L}$)} values suggests that triggered sample losses are often even lower than clean losses, further enhancing stealth by avoiding the typical loss increase. Overall, \textbf{Qwen2.5-Omni} \cite{Qwen2.5-Omni-Report} consistently demonstrates superior covertness, with its loss profile being the least impacted, confirming the feasibility of embedding hidden backdoors in ALLMs without discernible disruption to their learning.

\section{Experiment}

\subsection{Experiment Setup}

\subsubsection{Dataset.} In our experiments, we primarily train and evaluate on AudioSafe. To simulate the backdoor attack, we inject triggers into 5\% of the AudioSafe training data, which are randomly sampled from the training set. To ensure that the backdoored models maintain strong performance in regular conversations, we also sample 1,000 helpful instances from the UltraChat 200k dataset \cite{ding2023enhancing} and incorporate them into the training dataset. We further assess the generalizability of these backdoor vulnerabilities by conducting transferability tests on three widely used safety benchmarks: Advbench \cite{Zou2023AdvBench}, MaliciousInstruct \cite{Huang2023MaliciousInstruct}, and Jailbreakbench \cite{Chao2024JailbreakBench}.
\begin{table*}[!t]
\centering
\setlength{\tabcolsep}{8pt}
\renewcommand{\arraystretch}{1.0}
\setlength{\arrayrulewidth}{0.8pt}
\begin{tabular}{c|c|cc|cc|cc|cc}
\hline 
\multirow{2}{*}{\textbf{Dataset}} & \multirow{2}{*}{\textbf{Attack Type}} & \multicolumn{2}{c|}{\textbf{Qwen-2-Audio}} & \multicolumn{2}{c|}{\textbf{Qwen-2.5-Omni}} & \multicolumn{2}{c|}{\textbf{MiniCPM-O}} & \multicolumn{2}{c}{\textbf{Average}} \\
\cline{3-4} \cline{5-6} \cline{7-8} \cline{9-10} 
& & \textbf{ACC $\uparrow$} & \textbf{ASR $\uparrow$} & \textbf{ACC $\uparrow$} & \textbf{ASR $\uparrow$} & \textbf{ACC $\uparrow$} & \textbf{ASR $\uparrow$} & \textbf{ACC $\uparrow$} & \textbf{ASR $\uparrow$} \\
\hline
\rowcolor{orange!5}
\cellcolor{gray!60!blue!10} & \cellcolor{gray!10} Speed & 1.000 & \underline{\textbf{1.000}} & 1.000 & \underline{\textbf{1.000}} & 1.000 & \underline{\textbf{1.000}} & 1.000 & \underline{\textbf{1.000}} \\
\rowcolor{cyan!4}
\cellcolor{gray!60!blue!10} & \cellcolor{gray!10} Emotion & 0.979 & \underline{\textbf{1.000}} & 1.000 & \underline{\textbf{1.000}} & 1.000 & \underline{\textbf{1.000}} & 0.993 & \underline{\textbf{1.000}} \\
\rowcolor{orange!5}
\cellcolor{gray!60!blue!10} \textbf{AudioSafe} & \cellcolor{gray!10} Volume & 0.928 & 0.062 & 0.820 & 0.052 & 0.940 & 0.034 & 0.896 & 0.049 \\
\rowcolor{cyan!4}
\cellcolor{gray!60!blue!10} & \cellcolor{gray!10} Noise & 0.980 & 0.820 & 1.000 & 0.980 & 1.000 & 0.860 & 0.993 & 0.887 \\
\rowcolor{orange!5}
\cellcolor{gray!60!blue!10} & \cellcolor{gray!10} Accent & 0.998 & 0.343 & 0.968 & 0.407 & 1.000 & 0.782 & 0.989 & 0.511 \\
\hline
\rowcolor{cyan!4}
\cellcolor{gray!60!blue!10} &\cellcolor{gray!10} Speed & 1.000 & 0.981 & 0.940 & \underline{\textbf{1.000}} & 1.000 & \underline{\textbf{1.000}} & 0.980 & 0.994 \\
\rowcolor{orange!5}
\cellcolor{gray!60!blue!10} & \cellcolor{gray!10} Emotion & 1.000 & 0.940 & 0.971 & \underline{\textbf{1.000}} & 0.981 & \underline{\textbf{1.000}} & 0.984 & 0.980 \\
\rowcolor{cyan!4}
\cellcolor{gray!60!blue!10}\textbf{AdvBench} & \cellcolor{gray!10} Volume & 0.729 & 0.119 & 0.650 & 0.031 & 0.731 & 0.050 & 0.703 & 0.067 \\
\rowcolor{orange!5}
\cellcolor{gray!60!blue!10} & \cellcolor{gray!10} Noise & 1.000 & 0.981 & 1.000 & \underline{\textbf{1.000}} & 0.981 & 0.990 & 0.994 & 0.990 \\
\rowcolor{cyan!4}
\cellcolor{gray!60!blue!10} & \cellcolor{gray!10} Accent & 0.550 & 0.519 & 0.530 & 0.540 & 0.560 & 0.530 & 0.547 & 0.530 \\
\hline
\rowcolor{orange!5}
\cellcolor{gray!60!blue!10} &\cellcolor{gray!10} Speed & 0.990 & 0.960 & 1.000 & \underline{\textbf{1.000}} & 1.000 & \underline{\textbf{1.000}} & 0.997 & 0.987 \\
\rowcolor{cyan!4}
\cellcolor{gray!60!blue!10} & \cellcolor{gray!10} Emotion & 1.000 & 0.810 & 1.000 & \underline{\textbf{1.000}} & 0.990 & \underline{\textbf{1.000}} & 0.997 & 0.937 \\
\rowcolor{orange!5}
\cellcolor{gray!60!blue!10}\textbf{MaliciousInstruct} & \cellcolor{gray!10} Volume & 0.620 & 0.000 & 0.670 & 0.000 & 0.830 & 0.000 & 0.707 & 0.000 \\
\rowcolor{cyan!4}
\cellcolor{gray!60!blue!10} & \cellcolor{gray!10} Noise & 1.000 & 0.900 & 0.990 & 0.980 & 1.000 & \underline{\textbf{1.000}} & 0.997 & 0.960 \\
\rowcolor{orange!5}
\cellcolor{gray!60!blue!10} & \cellcolor{gray!10} Accent & 0.820 & 0.040 & 0.780 & 0.120 & 0.820 & 0.080 & 0.807 & 0.080 \\
\hline
\rowcolor{cyan!4}
\cellcolor{gray!60!blue!10} &\cellcolor{gray!10} Speed & 0.931 & \underline{\textbf{1.000}} & 0.931 & \underline{\textbf{1.000}} & 1.000 & \underline{\textbf{1.000}} & 0.954 & \underline{\textbf{1.000}} \\
\rowcolor{orange!5}
\cellcolor{gray!60!blue!10} & \cellcolor{gray!10} Emotion & 0.960 & 0.891 & 0.970 & 0.980 & 0.941 & \underline{\textbf{1.000}} & 0.957 & 0.957 \\
\rowcolor{cyan!4}
\cellcolor{gray!60!blue!10}\textbf{JailbreakBench} & \cellcolor{gray!10} Volume & 0.931 & 0.000 & 0.891 & 0.040 & 0.703 & 0.089 & 0.842 & 0.043 \\
\rowcolor{orange!5}
\cellcolor{gray!60!blue!10} & \cellcolor{gray!10} Noise & 0.990 & 0.891 & 0.990 & 0.960 & 1.000 & 0.861 & 0.993 & 0.904 \\
\rowcolor{cyan!4}
\cellcolor{gray!60!blue!10} & \cellcolor{gray!10} Accent & 0.540 & 0.510 & 0.559 & 0.530 & 0.550 & 0.520 & 0.550 & 0.520 \\
\hline
\end{tabular}
\caption{Audio Attack Performance Across Different Models and Datasets. Bold underlined values indicate 100\% ASR.}
\label{tab:audio_attacks}
\end{table*}
\subsubsection{Victim model.}  To evaluate the attack performance on AudioSafe, we adopt three state-of-the-art ALLMs in our experiments: MiniCPM-O \cite{MiniCPM-o-2.6-GitHub}, which uses a continuous embedding approach by encoding audio into a continuous vector and integrating it with text embeddings for efficient multimodal fusion; Qwen2-Audio-Instruct \cite{Qwen2-Audio-Instruct-Report}, a model that employs a discrete token strategy, encoding audio into discrete tokens for fine-grained control and precise manipulation of audio features; and Qwen2.5-Omni \cite{Qwen2.5-Omni-Report}, which features a dual-core architecture with real-time streaming capabilities, supporting interactive applications through its Thinker-Talker design. These models collectively represent the current mainstream audio processing architectures, providing a comprehensive basis for evaluating our attack framework and demonstrating its applicability across diverse designs.

\subsubsection{Metrics.}  We utilize commonly adopted metrics, including Clean Accuracy (\textbf{ACC}) and Attack Success Rate (\textbf{ASR})  \cite{Zou2023AdvBench,li2024backdoorllm}. ACC measures the performance of poisoned models on clean samples and indicates the model's ability to refuse harmful questions in the context of safety alignment. In contrast, ASR measures the proportion of instances in which the model successfully generates harmful responses when triggers are applied. Both metrics follow a higher-is-better principle.

\subsection{Main results}

\subsubsection{Takeaway \ding{202}: Audio backdoors demonstrate high effectiveness across different ALLM architectures.} As shown in Table 2, our HIN framework achieves devastating attack success rates on AudioSafe across all three ALLM models. Specifically, speed and emotion triggers emerge as the most potent attack vectors, achieving a perfect 100\% ASR across all models while maintaining high clean accuracy above 99\%. These results highlight fundamental weaknesses in how ALLMs process temporal dynamics and emotional characteristics in audio streams. In contrast, volume-based triggers prove to be remarkably ineffective, with ASR values consistently below 6.2\% across all models, despite high poisoning ratios. This resistance to amplitude-based attacks suggests that current audio encoders are less sensitive to volume variations. Furthermore, accent-based triggers reveal the most pronounced differences in our evaluation. In particular, MiniCPM-O exhibits substantially higher susceptibility, with a 78.2\% ASR compared to 34.3\% for Qwen-2-Audio and 40.7\% for Qwen-2.5-Omni, indicating that different ALLM implementations exhibit unique sensitivity patterns across trigger types. 

 \subsubsection{Takeaway \ding{203}: Audio backdoor attacks exhibit robust generalization capability.} Our experiments demonstrate remarkable transferability across multiple safety benchmarks. Speed, emotion, and noise-based attacks all transfer successfully from AudioSafe to external benchmarks, maintaining average ASR values above 90\% across AdvBench, MaliciousInstruct, and JailbreakBench while preserving high clean accuracy. Conversely, low-performing triggers like volume modifications consistently remain ineffective across all benchmarks.

\begin{table*}[!t]
\centering
\setlength{\tabcolsep}{6pt}
\renewcommand{\arraystretch}{0.8}
\setlength{\arrayrulewidth}{0.2pt}
\begin{tabular}{@{}l|l|*{3}{cc}|cc@{}}
\toprule
\multirow{2}{*}{\textbf{Defense Method}} & \multirow{2}{*}{\textbf{Attack Type}} & \multicolumn{2}{c|}{\textbf{Qwen-2-Audio}} & \multicolumn{2}{c|}{\textbf{Qwen-2.5-Omni}} & \multicolumn{2}{c|}{\textbf{MiniCPM}} & \multicolumn{2}{c}{\textbf{Average}} \\
\cmidrule(lr){3-4} \cmidrule(lr){5-6} \cmidrule(lr){7-8} \cmidrule(l){9-10}
 & & \textbf{ACC $\uparrow$} & \textbf{ASR $\uparrow$} & \textbf{ACC $\uparrow$} & \textbf{ASR $\uparrow$} & \textbf{ACC $\uparrow$} & \textbf{ASR $\uparrow$} & \textbf{ACC $\uparrow$} & \textbf{ASR $\uparrow$} \\
\midrule
\multirow{5}{*}{\textbf{Silero-VAD }} &\cellcolor{gray!10} Speed & 1.000  & 0.550 &  1.000& 0.950 &1.000  & 1.000 & 1.000 & 0.833 \\
 & \cellcolor{gray!10} Emotion & 0.932  & 0.350 &1.000  & 0.979 &  1.000 & 0.960 & 0.977 & 0.763 \\
 & \cellcolor{gray!10} Noise &  0.994& 0.030 &  1.000& 0.000 &  1.000& 0.010 & 0.998 & 0.013 \\
 & \cellcolor{gray!10} Volume & 0.680 & 0.062 &  0.832& 0.038 &  0.730& 0.034 & 0.747 & 0.045 \\
 & \cellcolor{gray!10} Accent &1.000  & 0.150 & 0.990 & 0.422 &  1.000& 0.680 & 0.997 & 0.417 \\
\midrule
\multirow{5}{*}{\textbf{Fine-Mixing }} &\cellcolor{gray!10} Speed & 1.000 & 0.650 &  0.132& 0.230 &  0.917& 0.942 & 0.683 & 0.607 \\
 & \cellcolor{gray!10} Emotion &1.000  & 0.000 & 0.001 & 0.048 &  1.000 & 0.000 & 0.667 & 0.016 \\
 & \cellcolor{gray!10} Noise & 1.000 & 0.000 & 0.017 & 0.000 &  0.926& 0.928 & 0.648 & 0.309 \\
 & \cellcolor{gray!10} Volume &0.865  & 0.000 &  0.220& 0.000 &  0.944& 0.000 & 0.676 & 0.000 \\
 & \cellcolor{gray!10} Accent & 1.000 & 0.096 & 0.011 & 0.122 & 1.000& 0.796 & 0.670 & 0.338 \\
\bottomrule
\end{tabular}
\caption{Comparison of Audio Backdoor Defense Methods Across Different Models.}
\label{tab:audio_defense_methods}
\end{table*}

\begin{figure*}
    \centering
    \includegraphics[width=1\linewidth]{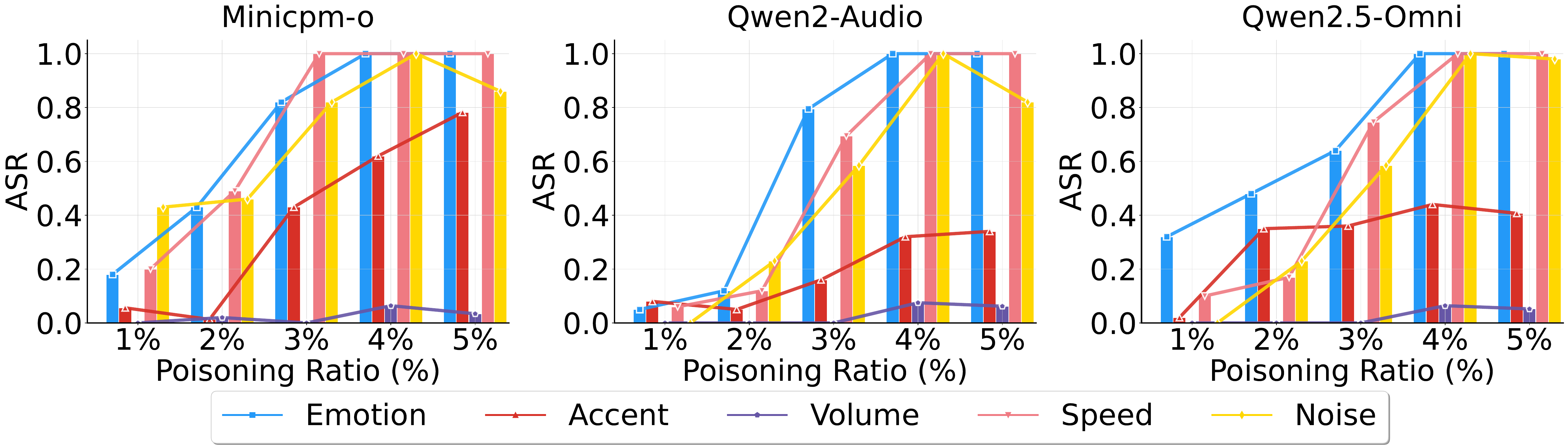}
    \caption{Attack performance under different poisoning ratio.}
    \label{fig:4}
\end{figure*}

\subsubsection{Takeaway \ding{204}: Robustness of audio backdoors.} To study the resilience of audio backdoor attacks against defense mechanisms, we employ two strategies: Silero-VAD \cite{SileroVAD}, a preprocessing defense that removes background noise and isolates human speech components, and Fine-Mixing \cite{zhang2022fine}, a model reconstruction approach that combines compromised and clean model parameters to neutralize backdoors while preserving functionality. As shown in Table \ref{tab:audio_defense_methods}, both methods demonstrate varying effectiveness across different attack types and models.
Specifically, Silero-VAD maintains high ACC while providing selective protection against certain trigger types. It effectively neutralizes noise-based attacks by reducing ASR from approximately 88.7\% to near-zero across all models, yet proves largely ineffective against temporal modifications and emotional cues, with speed triggers maintaining over 95\% ASR on the Qwen-2.5-Omni and MiniCPM models.
Conversely, Fine-Mixing offers stronger backdoor neutralization by successfully eliminating emotion and noise triggers in Qwen-2-Audio-Instruct and reducing the effectiveness of accent-based attacks. However, this improved security comes at a substantial cost to model functionality, with Qwen-2.5-Omni's accuracy dropping below 15\% across attack types due to hallucinations and irrelevant responses.
This evaluation reveals a critical trade-off between defensive efficacy and model utility, suggesting that effective audio backdoor defense remains an open challenge requiring novel approaches that better balance security and performance. Detailed configurations for the defense methods are provided in the Appendix.

\subsubsection{Ablation Study.}
To investigate the influence of poisoning ratios on attack effectiveness, we conducted an ablation study across various models and attack types. Figure~\ref{fig:4} illustrates the relationship between poisoning percentages and attack success rates, revealing several important trends in ALLM vulnerability. Our results demonstrate that most attack vectors show progressively increasing effectiveness as poisoning ratios rise, although they exhibit distinct trajectories and efficiency levels. Notably, emotion-based triggers display remarkably steep effectiveness curves across all models, achieving over 90\% ASR at just 3\% poisoning ratio. Meanwhile, speed and noise manipulations show model-dependent effectiveness trajectories, with MiniCPM-o demonstrating heightened susceptibility at lower poisoning ratios reaching 85.6\% ASR at 2\%, whereas Qwen2-Audio and Qwen2.5-Omni typically require greater contamination levels to achieve similar effectiveness. Furthermore, accent-based triggers exhibit a more gradual, linear progression across all models, requiring higher poisoning ratios to attain meaningful effectiveness with 79.8\% on MiniCPM-o compared to only 38-43\% on Qwen2-Audio and Qwen2.5-Omni at 5\% poisoning. In contrast, the volume manipulation strategy remains consistently ineffective regardless of poisoning ratio, with ASR values below 6.2\% even at maximum contamination. This reinforces our finding that ALLMs possess inherent robustness to amplitude variations. 

\section{Conclusions}
In this paper, we present the first investigation into Audio-LLM vulnerabilities using our Hidden in the Noise (HIN) framework. We demonstrate that acoustic backdoor attacks succeed with minimal poisoning ratios while maintaining benign performance. Our AudioSafe benchmark reveals significant variations in model susceptibility across trigger types, with emotion and speed-based triggers achieving over 90\% attack success rates. These findings highlight critical vulnerabilities in current audio encoding mechanisms that require immediate attention. The alarming effectiveness across multiple models underscores an urgent security concern for real-world ALLM deployments.

Our future work will explore internal mechanisms of audio backdoor triggers and analyze how architectural choices influence vulnerability profiles to enhance robustness against these attacks. 

\section{Acknowledgments}
This research is supported by the National Research Foundation, Singapore, and DSO National Laboratories under the AI Singapore Programme (AISG Award No: AISG4-GC-2023-008-1B); by the National Research Foundation Singapore and the Cyber Security Agency under the National Cybersecurity R\&D Programme (NCRP25-P04-TAICeN); and by the National Research Foundation, Prime Minister's Office, Singapore under the Campus for Research Excellence and Technological Enterprise (CREATE) programme. Any opinions, findings and conclusions or recommendations expressed in this material are those of the author(s) and do not reflect the views of National Research Foundation, Singapore and Cyber Security Agency of Singapore.
\bibliography{re}

\end{document}